\documentclass[12pt]{article}
\usepackage{times}

\include{psfig}
\usepackage{epsfig}
\begin{document}

\title{Associative Recall in Non-Randomly Diluted Neuronal Networks}

\author{ Luciano da Fontoura Costa \\
 Cybernetic Vision Research Group \\
 Instituto de F\'{\i}sica de S\~ao Carlos \\
 University of S\~ao Paulo \\
 13560-970  S\~ao Carlos, SP, Brazil \\
 \em{luciano@if.sc.usp.br}\\
\medskip\\
Dietrich Stauffer \\
Institute for Theoretical Physics \\
Cologne University \\
D-50923 K\"oln, Euroland \\
\em{stauffer@thp.uni-koeln.de}
}

\maketitle

\abstract{The potential for associative recall of diluted neuronal
networks is investigated with respect to several biologically relevant
configurations, more specifically the position of the cells along the
input space and the spatial distribution of their connections.  First
we put the asymmetric Hopfield model onto a scale-free
Barab\'asi-Albert network. Then, a geometrical diluted 
architecture, which maps from $L$-bit input patterns into $N$-neurons
networks, with $R=N/L<1$ (we adopt $R$=0.1, 0.2 and 0.3), is
considered.  The distribution of the connections between cells along
the one-dimensional input space follows a normal distribution centered
at each cell, in the sense that cells that are closer to each other
have increased probability to interconnect. The models also explicitly
consider the placement of the neuronal cells along the input space in
such a way that denser regions of that space tend to become denser,
therefore implementing a special case of the Barab\'asi-Albert
connecting scheme.  The obtained results indicate that, for the case
of the considered stimuli and noise, the network performance increases
with the spatial uniformity of cell distribution.}

\section{Introduction}

A number of mathematical concepts and tools have been extensively
applied in order to construct models and simulations of neuronal
behavior that could shine some light into the intricate workings of
the brain.  In physics, particular importance and interest have been
placed on Hopfield models because of their potential for associative
recall and close relationship with statistical mechanics
\cite{hopfield}.  However, while most such approaches have considered
fully connected structures, experimental evidences indicate that
biological neuronal networks involve a combination of intense local
interconnections (responsible, among other things, for lateral
inhibition \cite{kandel}) coexisting with long range mappings which
often exhibit topographical structure, in the sense that the
connections preserve the spatial adjacency of the signal.  Studies
addressing partially connected Hopfield networks on lattices have
appeared sporadically in the literature \cite{kurten}.  More recently
\cite{stauffer}, the Barab\'asi-Albert \cite{ba} connecting scheme,
where the ``rich gets richer'', was used to define Hopfield
architectures with symmetric couplings whose potential for associative
recall was assessed with respect to the number of connections between
the constituent cells.  The obtained results indicated that, despite
the relatively sparse and heterogeneous connections, for suitable
parameters such architectures preserved a good deal of the associative
recall when compared to fully connected counterparts, bridging the gap
to further investigations.

The present work reports the extension of the models described in
\cite{Costa,stauffer} in three aspects that are specially relevant
from the biological point of view, namely the asymmetry that neuron
$i$ may influence neuron $k$ even if neuron $k$ has no influence on
neuron $i$, the spatial distribution of cells along the input space
(assumed to be a one-dimensional vector), and the spatial distribution
of the connections between those cells. More especifically, as the
network is constructed, each added cell is placed preferentially at
regions of the input space that are denser in cells, in such a way
that dense regions tend to become denser.  Being diluted, not every
bit of the input pattern will be connected to a neuron.  Also, the
connections between a cell $i$ and the rest of the network are
established in such a way that cells that are closer to each other
have higher probability to interconnect, which is achieved through
Monte Carlo simulation assuming normal weighting centered at cell $i$.
The potential of the networks for associative recall is experimentally
assessed with respect to several parameter configurations, with
special attention given to the effect of the uniformity of the cell
spatial distribution over the obtained performance.

The article starts by revising some key aspects of biological neuronal
structures and proceeds by presenting the obtained results and
respective discussion, with special attention placed on the biological
interpretation and implication of the observed behavior in terms of
the parameter configurations.

\section{Spatial Organization in Biological and Computational Networks}

First, in what we call the asymmetric Hopfield-Barab\'asi-Albert
model, we put the Hopfield model onto the scale-free Barab\'asi-Albert
\cite{ba} network.  In contrast to spin glasses and many physics
systems, the interactions between biological neurons are not
symmetric: Neuron $k$ may influence neuron $i$ even if neuron $i$ has
no influence on neuron $k$. In other words, the synaptic strengths
$J(i,k)$ are not symmetric, i.e.  $J(i,k)$ is not equal to
$J(k,i)$. We thus construct our model in such a way that every neuron
$i$ selects itself plus exactly $m$ other neurons (repeated
connections are allowed) as neighbours.  This fixed number $m+1$ of
influencing neighbours greatly simplifies the algorithm.

Second, in what we call the geometrical model, we try to incorporate
the geometry instead of only the topology of neural nets.  Another
characteristic feature exhibited by the mammalian cortex is the
spatial structuring permeating both the input signal and neuronal
architectures.  For instance, the visual field is mapped onto the
retina space, which innervates, through the lateral geniculate
nucleus, into the primary visual cortex in such a way that the spatial
adjancency of the input patterns is maintained.  The importance of
such \emph{topographical mappings} is corroborated by the fact that
they are found ubiquitously along the cortex, and not only in its
primary sensory regions.  It should be observed that such mappings are
not isometric. One important neuronal pheonomenon closely related to
topographical mappings is known as \emph{lateral inhibition}, where
the activity at a given cell is inhibited by its spatial neighbours
along the input space.  In spite of its key role in cortical
organization, relatively few neuronal networks have explicitly taken
spatial organization into account, Kohonen's SOM model \cite{kohonen}
being one of the pioneering exceptions.  At the same time, much of the
efforts devoted to Hopfield models have systematically overlooked the
spatial structure of input and topographical mappings.  The situation
is similar in the biological community, where only more recently
(e.g. \cite{manoel}) attention was focused on this important
organizational principle.  The fact that neuronal cells that are
closer together tend to have increased chance to interconnect
(e.g. \cite{kandel}) has also been often overlooked.  Indeed, cortical
organization seems to combine local, intensive connections with long
distance, sparser, mappings between distinct neuronal modules.
Such local/global organization may also underly other spatial scales
structures in the brain, possibly leading to scale-free
characteristics over limited spatial scale intervals.  In the current
work, we propose a novel variation of the Hopfield model for
associative recall that explicitly takes into account the spatial
organization of the neuronal cells and the localized and global
spatial distribution of their connections.

The input space is assumed to be the one-dimensional vector $I(i)$,
$i=1, 2, \ldots,$ $L$, to which a total of $N$ neuronal cells are
attached.  The number of connections made by each cell is fixed and
equal to $m$.  The network is diluted, in the sense that $R=N/L<1$.
Figure 1 illustrates a typical realization of the assumed neuronal
architecture for $N=7$, $m=3$ and $L=15$, with the respective
asymmetric weight matrix shown in Equation (~\ref{eq:wm}).  Each line
$i$ shows the connections of neuron $i$ to other neurons by a '1'.
The construction of the network involves the two following steps: (a)
neurons are placed at spatial positions $p \in {0, 1, \ldots, L}$
according to a probability density function $h(p)$ (cells can only be
added at empty sites); and (b) $m$ not necessarily distinct synaptical
connections are established between the just added cell and its
neighbours according to a normal (Gaussian) density function $g_\sigma
(d)$, where $d$ is the distance between the position $p$ of the
reference cell and each of the other cells in the network and $\sigma$
is the standard deviation of the normal distribution.  It is assumed
throughout this article that $L=1000$.

\begin{figure}[ht]  
\begin{center}
  \epsfig{file=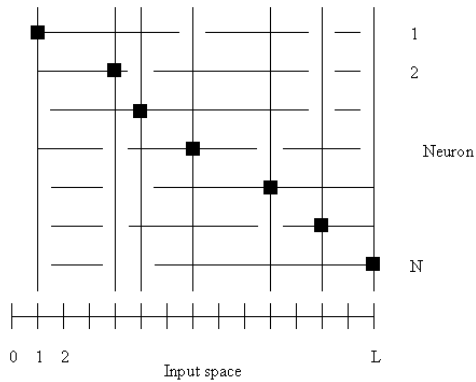, width=0.4\textwidth} \\ 
  \caption{Example of
    the networks considered in this article assuming $N=7$, $m=3$ and
    $L=15$. The black squares stand for the position of the neurons.}
  \end{center}
\end{figure}

\begin{equation}   \label{eq:wm}
  J = \left [  \begin{array}{ccccccc}
                 1 & 1 & 1 & 0 & 1 & 0 & 0 \\   
                 1 & 1 & 0 & 1 & 1 & 0 & 0 \\   
                 0 & 1 & 1 & 1 & 1 & 0 & 0 \\   
                 1 & 0 & 1 & 1 & 0 & 1 & 0 \\   
                 0 & 0 & 0 & 1 & 1 & 1 & 1 \\   
                 0 & 0 & 1 & 1 & 0 & 1 & 1 \\   
                 0 & 0 & 0 & 1 & 1 & 1 & 1 \\   
               \end{array}  \right] \\
\end{equation}

A non-stationary (along time) density function $h(p)$ is obtained at
each interaction during the network construction by adding the
function $f(i)=1$ if $|i-p| \leq a$ and $0$ otherwise to the previous
instance of $h(p)$, where $p$ is the position of the most recently
added cell, and renormalizing $h(p)$.  Standard Monte Carlo sampling
is then performed over the respective distribution function so as to
select the position $p$ for a new cell.  The probability function
$h(p)$ is initially set as uniform.  Monte Carlo simulation is also
employed to select $m$ cells for symmetric connections with the
current cell.  The connections are performed as follows: (i) a vector
$v(i)$ is built such that $v(i)=1$ at the positions $i$ where neurons
exist and $0$ otherwise; (ii) this vector is weighted according to a
normal density function centered at $p$, the position of the current
cell; and (iii) Monte Carlo is used to select $m$ sites of $v$.  It is
observed that all cells are always self-connected.  Figure 2
illustrates typically obtained connection matrices.  The reinforced
diagonal structure of such matrices favors conditioning and reflects
the locality of the synapses at specific spatial scales defined by
$\sigma$.

\begin{figure}[ht] 
  \begin{center}
    \begin{tabular}{cc}
     \epsfig{file=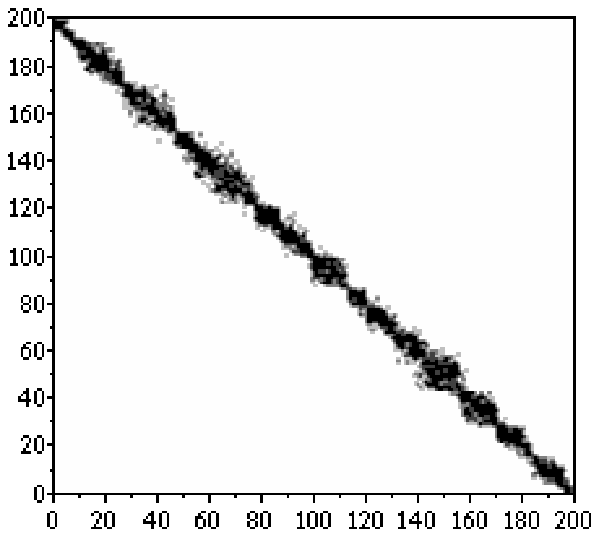, width=0.4\textwidth} (a) 
     \epsfig{file=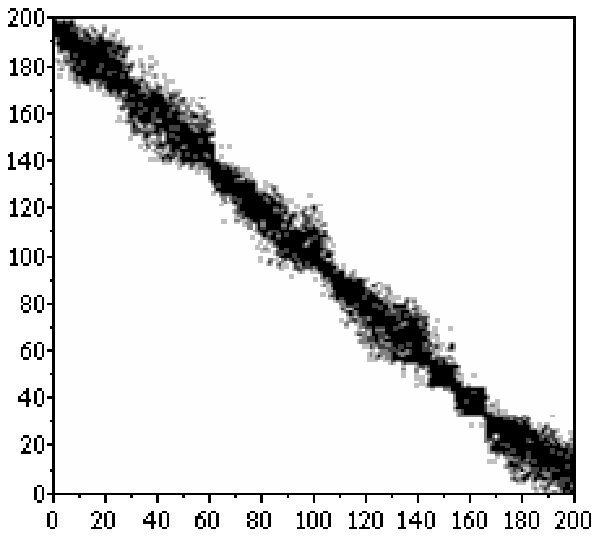, width=0.4\textwidth} (b) \\ 
     \epsfig{file=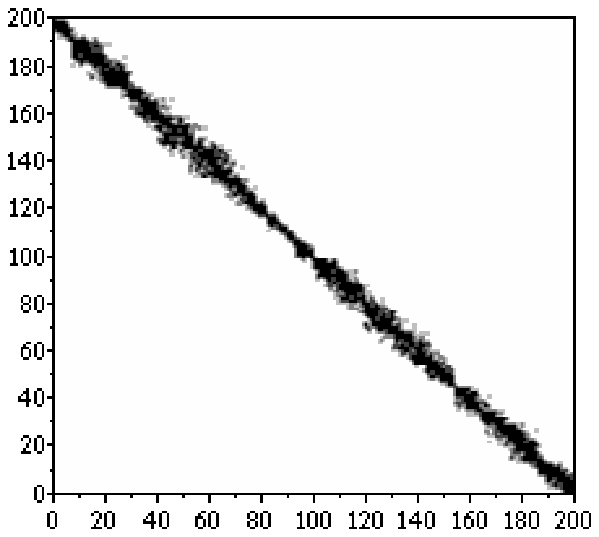, width=0.4\textwidth} (c)  
     \epsfig{file=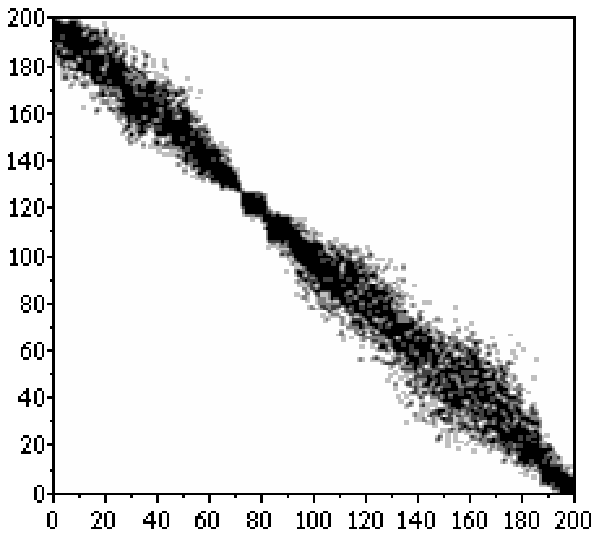, width=0.4\textwidth} (d) \\ 
    \end{tabular}
    \caption{Example of connection matrices, shown as gray-color images,
    for $m=25$ and $a= 10$ (a) and $100$ (c), and for $m=50$ and
    $a= 10$ (b) and $100$ (d).  Black means unconnected
    neurons, white means connected neurons.}
  \end{center}
\end{figure}

For both the Hopfield-Barab\'asi-Albert and the geometrical model, 
the neurons $i, \; i = 1,2, \dots N,$ in our models are either firing
($S_i =1$) or silent ($S_i =-1$) and are updated according to $$S_i
\rightarrow {\rm sign} (\sum_k J_{ik} S_k)$$ with synaptic strengths
$J_{ik} = \sum_\mu \xi_i^\mu \xi_k^\mu$ (Hebb rule) if $i$ and $k$ are
connected. Here $\xi_i^\mu = \pm 1, \; \mu = 1,2, \dots P,$ are $P$
random bit-strings called patterns, and one of them is supposed to be
recalled by this updating rule if it is presented to it in a perturbed
fashion: Associative memory. The quality of recall is measured by the
overlap $\Psi = \sum_i S_i \xi_i^1$ if, without loss of generality, the
first pattern is supposed to be recovered.

\section{Results}

For the asymmetric Hopfield-Barab\'asi-Albert model, $N$ binary
neurons are added to the initial core of $m+1$ neurons, and $P$ binary
patterns are stored. Initially, the first of these patterns is
presented to the network with ten percent of the sites flipped
(overlap = 0.8), and after a few iterations the dynamics comes to a
fixed point for which the overlap $\Psi$ is determined. Fig. 3 shows
as in the symmetric case that for $P \ll m \ll N$ the desired pattern
is fully restored. However, Fig. 4 shows that the scaling law: $\Psi =
f(m/P)$ for infinite $N$ is not fulfilled well, in contrast to the
symmetric case. The distribution of the number of neurons, influencing
$q$ sites each, follows roughly a $1/q^3$ law for small $q$ but not
for large $q$ (not shown).

\begin{figure}[ht] 
  \begin{center}
    \epsfig{file=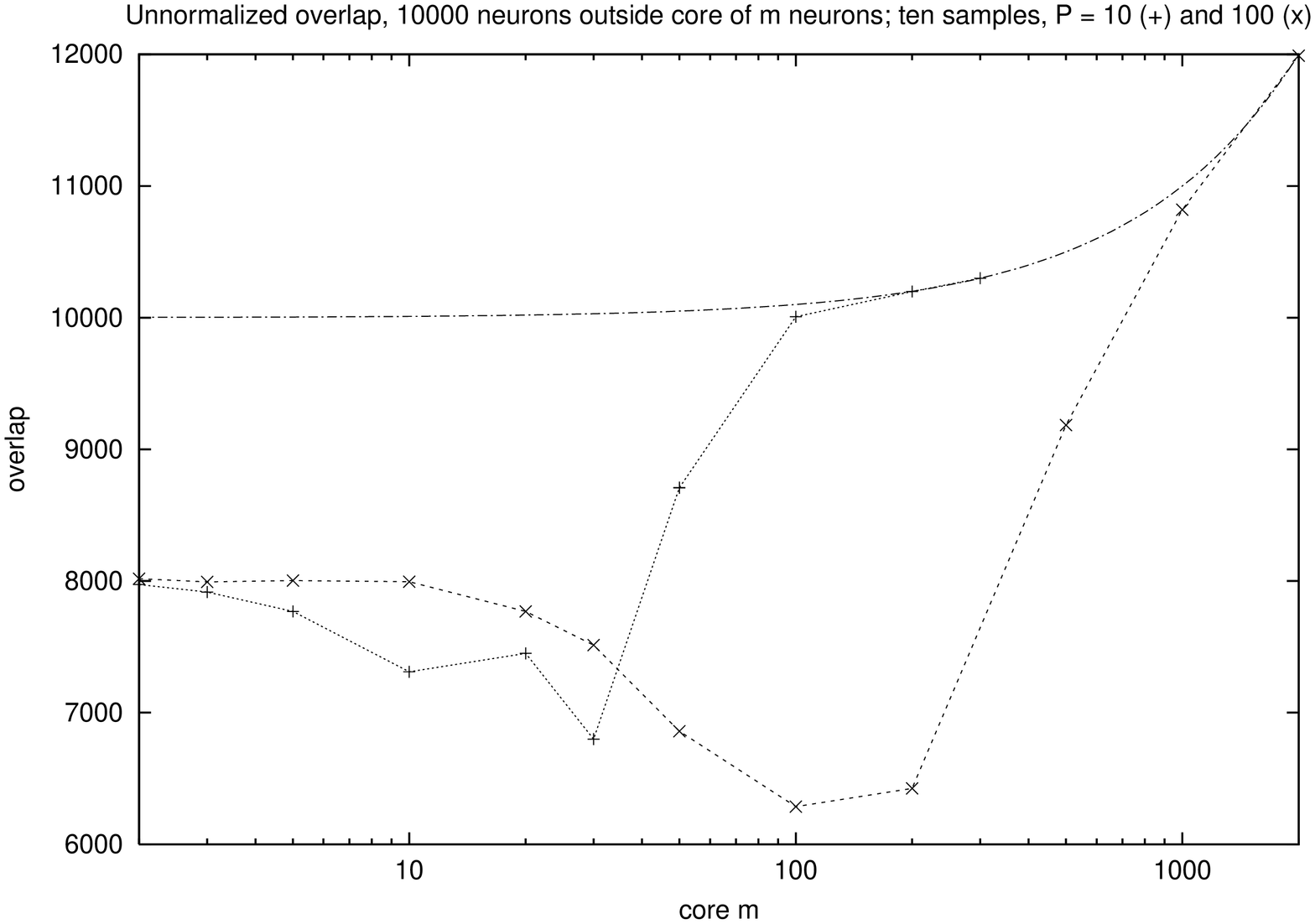, angle=-90, width=0.7\textwidth} 
   \caption{$\Psi$ for $10000 = m$ neurons ($m$ in the core, surrounded by
     10,000 others), for 10 (left) and 100 (right) patterns. The curve gives the
     overlap for full recovery.}
  \end{center}
\end{figure}

\begin{figure}[ht]  
  \begin{center}
    \epsfig{file=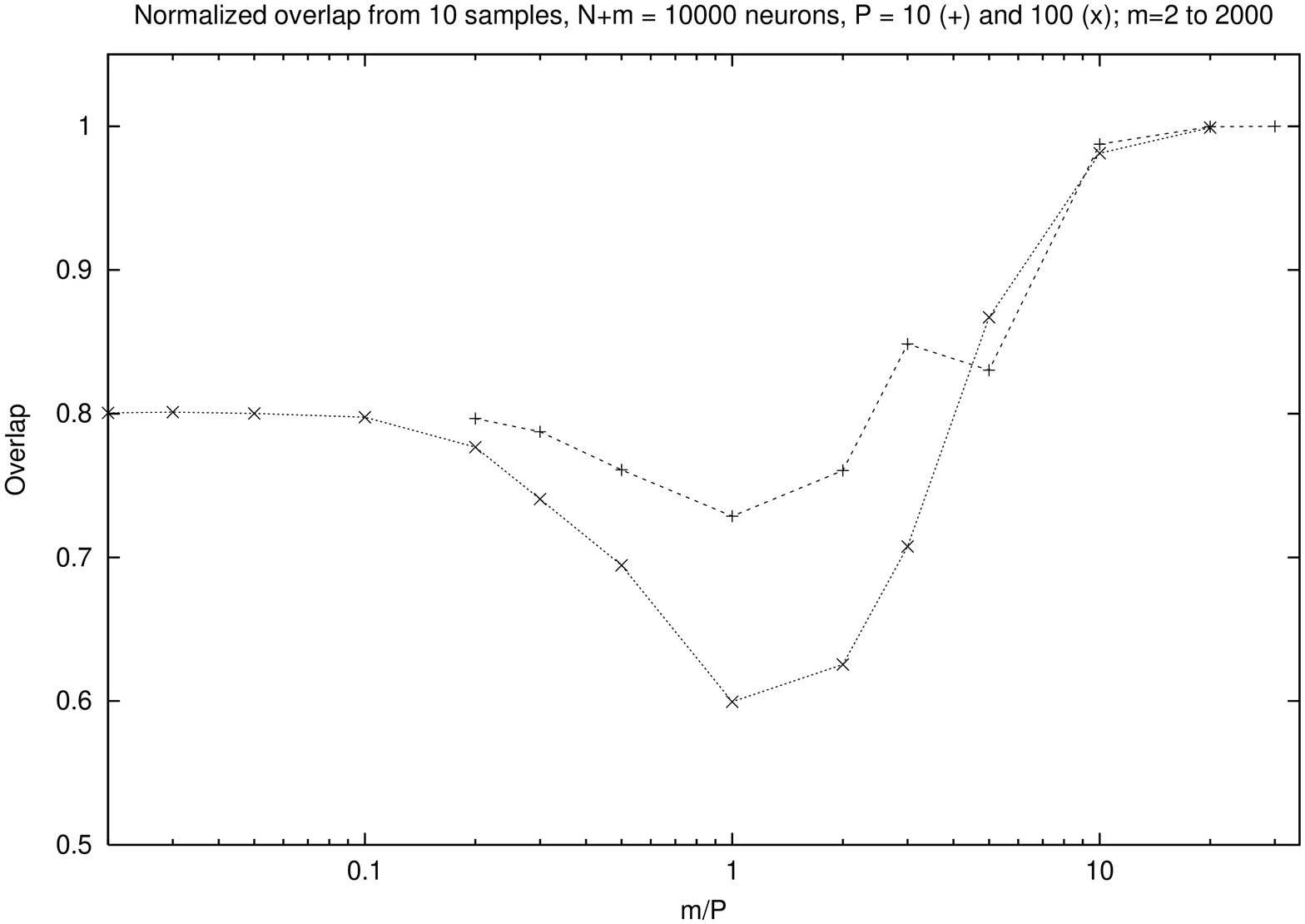, angle=-90,width=0.7\textwidth} 
   \caption{$\Psi/10000$ for $m$ neurons in the core and $N=10000-m$ neurons
     surrounding them, for 100 and 1000 patterns. If scaling would be valid
     the two data sets would overlap in this plot.}
  \end{center}
\end{figure}

For the geometrical model, simulations were performed considering
combinations of $m=25$ and $50$ with $a=10$ and $100$.
Parallel updating is adopted for Hopfield recovery and the number of
trained patterns is $P=0.05N$.  Relative overlaps are used ,
i.e. $R=\Psi/N$. The relative overlaps are calculated taking into
account only the neurons attached to the input pattern.

Figure 5 shows the correlation integral $C(r)$ \cite{schroeder},
namely the number of neurons with distance along the input space
smaller or equal to $r$ divided by $N^2$. Figures 5 to 7 show examples
of typically obtained density and respective distribution functions
obtained for $a=10$ (a) and $100$ (b) and considering 10 different
realizations of the spatial distribution of cells for each parameter
settings.

\begin{figure}[ht]
  \begin{center}
    \epsfig{file=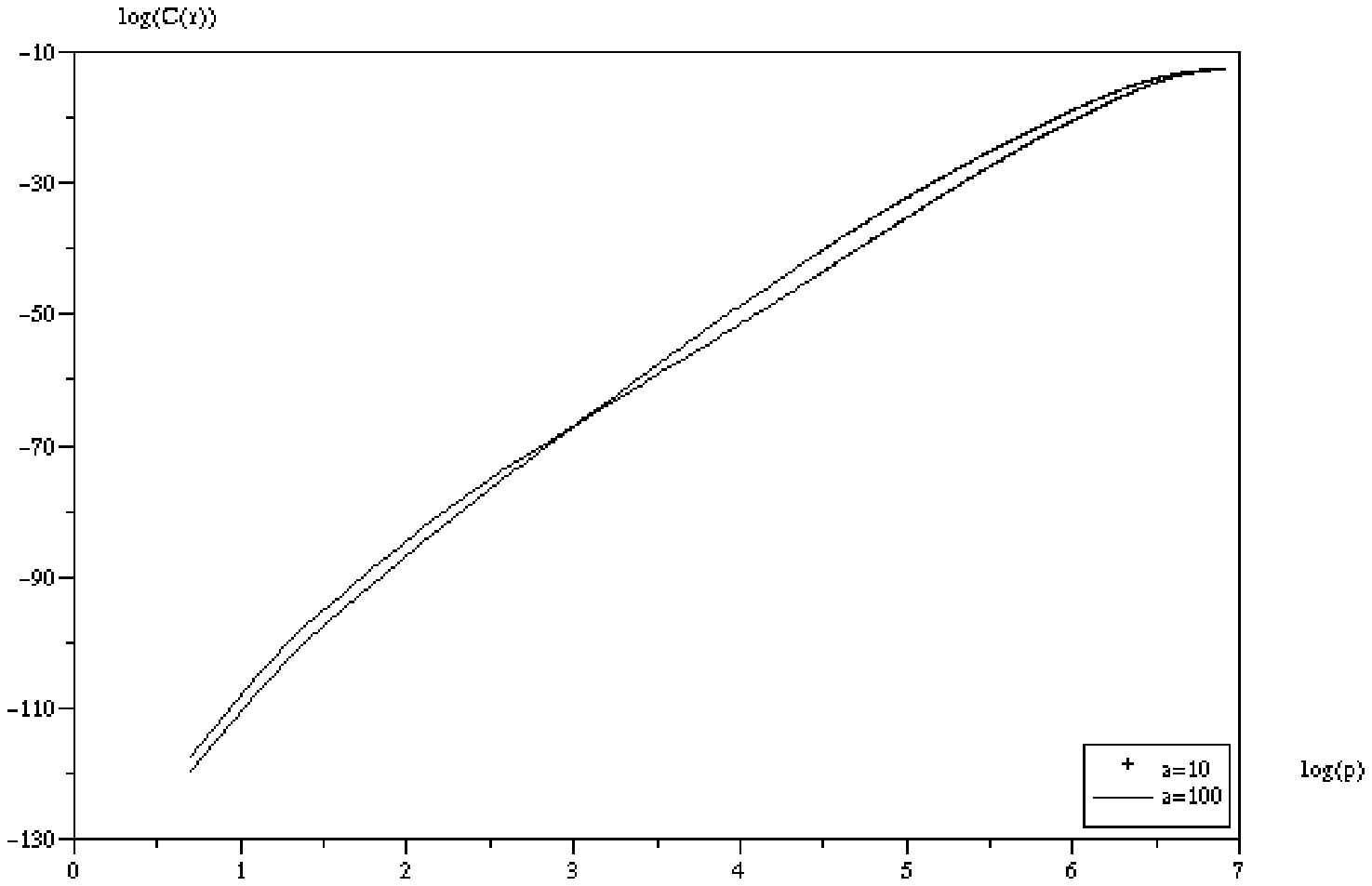, width=0.5\textwidth} \\ 
    \caption{The loglog
    graph of the correlation integral $C(r)$ obtained for $a=10$ (a) and
    $100$ (b).}
  \end{center}
\end{figure}

\begin{figure}[ht]  
  \begin{center}
    \begin{tabular}{cc}
    \end{tabular}
  \caption{Typically obtained density (up) and respective distribution
   functions (down) obtained for $a=10$ and $100$ (b).}
  \end{center}
\end{figure}

\begin{figure}[ht] 
  \begin{center}
    \caption{Relative overlaps for $m=25$ and $a=10$ and $100$ (upper pair
    of curves) and $m=50$ and $a=10$ and $100$ (lower pair).}
  \end{center}
\end{figure}

\section{Discussion}

For the Hopfield-Barab\'asi-Albert model, the asymmetric version gave
results similar to the symmetric version \cite{stauffer}, provided
self-interaction in taken into account in both cases.  For the
geometrical model, as indicated by Fig. 5, the considered scheme for
positioning the neurons along the input space led to similar power law
behavior for $a=10$ and $100$ for the considered correlation lags.  At
the same time, as is clear from Fig. 6, the value of $a$
strongly influences the characteristic spatial distribution of
neurons, in the sense that shorter clusters of cells, indicated by the
clumps in Fig. 6, are obtained for smaller values of $a$.
As expected, higher values of $m$ tend to produce more widespread
distribution of synaptic connections, leading to the wider dispersions
around the main diagonal of the matrices in Fig. 2(b) and (d) and
substantially superior potential for associative recall, shown in
Fig. 7.  This is a consequence of the fact that a larger number of
connections has direct impact over the amount of memory in the
network.  At the same time, the network performance tends to improve
as $a$ is increased (see Fig. 7).  A more uniform spatial distribution
of the neurons is obtained for larger values of $a$, with the limit
case $a \rightarrow \infty$ tending to the uniform distribution. It is
clear that, at least for the considered kind of stimuli and noise
(uniform distribution of signal changes), the effect of irregular
distribution of neurons along the input pattern space tends to
deteriorate associative recall.  The important biological implication
is that the development (ontogeny) of real neuronal systems coping
with uniformly distributed stimuli and noise has to incorporate
mechanisms for ensuring uniform distribution of neuronal placement.
In addition to topographical maps, other possible biological
mechanisms that could be involved in such spatial tuning may involve
controled apoptosis (i.e. programmed neuronal cell death) and
chemotatic migration modulated by density.  It is important to bear in
mind that different results could be obtained for distinct input
stimuli and noise models.  For instance, stimuli characterized by
specific correlation lengths could be better processed by networks
with neurons and connections presenting congruent spatial
characteristis.

All in all, the present work has considered extensions of the
classical Hopfield network which are characterized by enhanced
biological plausibility as far as asymetric connections and spatial
distribution of cell bodies and synapses are concerned. Such models
were analysed according to the shape-function paradigm, i.e. by
assessing how the model parameters influenced the associative recal
performance, and a series of perspectives has been opened for further
investigation.  A particularly interesting issue being currently
addressed regards the characterization of the effect of the individual
morphology of real neuronal cells over the network behavior, as well
as the influence of stimuli and noise spatial features over the
respective network performance.

\paragraph{Acknowledgments} 

Luciano da F. Costa is grateful to the Human Frontiers Science
Program, FAPESP (99/12765-2) and CNPq (468413/00-6 and 301422/92-3)
for financial help. D. Stauffer thanks H. Sompolinsky for suggesting
asymmetric networks and GIF for travel support.


\begin{thebibliography}{99}

\bibitem{hopfield} J.J. Hopfield, Proc. Natl. Acad. Sci. USA {\bf 79}, 2554 
(1982).

\bibitem{kandel} E. R. Kandel, J. H. Schwartz and T. M. Jessel,
Principles of Neural Science, Appleton and Lange, 1991.

\bibitem {kurten} K.E. K\"urten, J. Physique {\bf 51}, 1585 (1990);
 B.M. Forrest, J. Physique {\bf 50}, 2003 (1989);
D. Ji, B. Hu and T. Chen, Physica A 229, 147 (1996). See also:
O. Shefi et al., Phys, Rev. E 
{\bf 66}, 021905 (2002); S. Morita et al., Physica A {\bf 298}, 553 (2001);
J. Karbowski, Phys. Rev. Lett. {\bf 86}, 3674 (2001).

\bibitem {stauffer} D. Stauffer, A. Aharony, L. da F. Costa and J. Adler, 
preprint cond-mat/0212601.

\bibitem{ba} A.L. Barab\'asi and R. Albert, Science {\bf 286}, 509 (1999);
 R. Albert and A.L. Barab\'asi, Rev. Mod. Phys. {\bf 74}, 47 (2002);
 S.N. Dorogovtsev and J.F.F. Mendes, Adv. Phys. {\bf 51}, 1079 (2002).

\bibitem{Costa} L. da F. Costa and E. T. M. Manoel,
Neuroinformatics {\bf 1}, 66 (2002).

\bibitem{kohonen} T. Kohonen, Self-Organizing Maps, Springer-Verlag
(2001).

\bibitem{manoel} L. da F. Costa, E. T. M. Manoel, F. Faucereau,
J. Chelly, J. van Pelt and G. Ramakers, Network {\bf 13}, 283 (2002);
G. Ascoli, Computational  Neuroanatomy: Principles and Methods, 
Humana Press (2002).

\bibitem{schroeder} M. Schroeder, Fractal, Chaos, Poser Laws,
W. H. Freeman (1991).


\end{thebibliography}
\end{document}